\documentclass{PoS}
\usepackage{amssymb,amsmath,graphicx,epsfig,bm,color}
\usepackage{bm}
\newcommand{\be}{\begin{equation}}
\newcommand{\ee}{\end{equation}}
\newcommand{\bea}{\begin{eqnarray}}
\newcommand{\eea}{\end{eqnarray}}
\renewcommand{\d}{\mathrm{d}}
\newcommand{\pup}{p^\uparrow}

\title{Process dependence of the gluon Sivers function in inclusive $pp$ collisions: phenomenology}

\ShortTitle{Process dependence of the gluon Sivers function}

\author{\speaker{Umberto D'Alesio}\\
        Dipartimento di Fisica, Universit\`a di Cagliari  and
        INFN, Sezione di Cagliari, Cittadella Universitaria, I-09042 Monserrato (CA), Italy\\
        E-mail: \email{umberto.dalesio@ca.infn.it}}

\author{Carlo Flore\\
        Dipartimento di Fisica, Universit\`a di Cagliari  and
        INFN, Sezione di Cagliari, Cittadella Universitaria, I-09042 Monserrato (CA), Italy\\
        E-mail: \email{carlo.flore@ca.infn.it}}

\author{Francesco Murgia\\
        INFN, Sezione di Cagliari, I-09042 Monserrato (CA), Italy\\
        E-mail: \email{francesco.murgia@ca.infn.it}}

\author{Cristian Pisano\\
        Dipartimento di Fisica, Universit\`a di Cagliari  and
        INFN, Sezione di Cagliari, Cittadella Universitaria, I-09042 Monserrato (CA), Italy\\
        E-mail: \email{cristian.pisano@ca.infn.it}}

\author{Pieter Taels\\
        INFN, Sezione di Cagliari, I-09042 Monserrato (CA), Italy\\
        E-mail: \email{pieter.taels@ca.infn.it}}

\abstract{Within the so-called color gauge invariant generalized parton model, 
a TMD scheme including initial- (ISI) and final-state (FSI) interactions, 
we present a phenomenological analysis of available SSA data for pion and $D$-meson production in $pp$ collisions. 
This allows us, for the first time, to put a preliminary constraint on the two universal types of gluon 
Sivers function entering the model. 
Predictions for SSAs in $J/\psi$ and direct photon production, as well as a comparison with the simpler 
generalized parton model (without ISIs and FSIs), are also presented.}

\FullConference{23rd International Spin Physics Symposium - SPIN2018 -\\
		10-14 September, 2018\\
		Ferrara, Italy}

\begin{document}

\section{Introduction and formalism}

Transverse momentum dependent parton distributions (TMDs) are nowadays believed to represent fundamental pieces in our  
comprehension of the internal structure of the nucleons. 
They are also crucial to explain very interesting spin-polarization phenomena like 
single spin asymmetries (SSAs). The Sivers function~\cite{Sivers:1989cc}, 
related to the asymmetry in the azimuthal distribution of unpolarized partons 
inside a high-energy transversely polarized proton, is of great interest. 
Its existence is well established and many phenomenological extractions of its quark component are available. 
In contrast, the gluon Sivers function (GSF) is much more difficult to access and still poorly known.

In Ref.~\cite{DAlesio:2015fwo} a first extraction of the GSF from mid-rapidity data on 
SSAs in $p^\uparrow p \to \pi^0\,X $~\cite{Adare:2013ekj} 
has been obtained within a TMD approach, the so-called generalized parton model (GPM). 
In this scheme, applied to single-scale processes, TMDs are taken to be universal. 
Although not formally proven, it has been shown to be phenomenologically very successful 
(see Refs.~\cite{DAlesio:2007bjf, Aschenauer:2015ndk}).

A QCD extension of this model, the Color Gauge Invariant GPM (CGI-GPM), 
was developed in Refs.~\cite{Gamberg:2010tj,DAlesio:2011kkm}, 
focusing on the process dependence of quark Sivers function. 
Here, the effects of initial- (ISI) and final- (FSI) state interactions 
are taken into account, within a one-gluon exchange approximation. 
In particular, the CGI-GPM can reproduce the opposite relative sign of the quark Sivers 
functions in SIDIS and in the Drell-Yan processes~\cite{Collins:2002kn,Brodsky:2002cx}. 

In Ref.~\cite{DAlesio:2017rzj} the CGI-GPM has been extended, for the first time, to the 
gluon Sivers effect in SSAs for inclusive $J/\psi$ and $D$-meson production 
in $pp$ collisions, while inclusive pion and photon production 
has been considered in Ref.~\cite{DAlesio:2018rnv}. 
In this scheme, two different classes of modified partonic cross sections enter, 
each convoluted with a different GSF. These universal and independent distributions 
are named the $f$-type and $d$-type gluon Sivers functions.

We present here a phenomenological analysis, giving only few basic expressions      
and referring to Refs.~\cite{DAlesio:2017rzj,DAlesio:2018rnv} for all details. 
The SSA for the inclusive process $p^\uparrow p \to h \, X$, defined as
\begin{equation}
A_N \equiv \frac{\d\sigma^\uparrow-\d\sigma^\downarrow}{\d\sigma^\uparrow+\d\sigma^\downarrow} 
\equiv\, \frac{ \d\Delta\sigma}{ 2 \d\sigma}\,,
\end{equation}
within a TMD scheme and in suitable kinematical regions, is mainly sensitive to the Sivers function:
\be
\Delta \hat f_{a/\pup}\,(x_a, \bm{k}_{\perp a})
\label{defsiv}
= \Delta^N f_{a/\pup}\,(x_a, k_{\perp a}) \> \cos\phi_a
=  -2 \, \frac{k_{\perp a}}{M_p} \, f_{1T}^{\perp a} (x_a, k_{\perp a}) \> \cos\phi_a \, .
\ee
In particular, for $D$-meson, $J/\psi$ or mid-rapidity $\pi\;(\gamma)$ production 
in $pp$ collisions only the gluon Sivers effect can be sizeable, 
being all these processes dominated by gluon initiated sub-processes.

Within the CGI-GPM the $f$- and $d$-type contributions to $\d\Delta\sigma$, \emph{to be added up}, read
\begin{align}
  \left ( -\frac{k_{\perp g}}{M_p} \right ) f^{\perp g\, (f,d)}_{1 T}(x_g, k_{\perp g})\cos\phi_g \otimes
\> f_{b/p}(x_b, \bm{k}_{\perp b}) \otimes H^{{\rm Inc}\, (f,d)}_{gb \to cd}
\otimes \>  D_{h/c}(z, \bm{k}_{\perp h}) \>,
\label{sivgen}
\end{align}
where $H^{\rm Inc}$ include ISIs and FSIs. 
For $\gamma$ production $D_{h/c}$ is replaced by a Dirac delta-function on $z$ and $\bm{k}_{\perp h}$, 
and for $J/\psi$ production by a proper non perturbative factor. 
In the GPM approach only one GSF appears and $H^{\rm inc}$ is replaced 
by the ordinary unpolarized partonic contribution $H^U$.

The first $k_\perp$-moment of the Sivers function is also of relevance:
\be
\Delta^N \! f_{g/p^\uparrow}^{(1)}(x) = 
\int \d^2 \bm{k}_\perp \frac{k_\perp}{4 M_p} \Delta^N \! f_{g/\pup}(x,k_\perp) \equiv - f_{1T}^{\perp (1) g}(x) \, .
\label{siversm1}
\ee
For the gluon Sivers function we adopt a factorized Gaussian-like form:
\begin{equation}
\Delta^N\! f_{g/p^\uparrow}(x,k_\perp) =   
\left (-2\frac{k_\perp}{M_p}  \right )f_{1T}^{\perp\,g} (x,k_\perp)  = 2 \, {\cal N}_g(x)\,f_{g/p}(x)\,
h(k_\perp) \,\frac{e^{-k_\perp^2/\langle k_\perp^2 \rangle_g}}
{\pi \langle k_\perp^2 \rangle_g}\,,
\label{eq:siv-par-1}
\end{equation}
where $f_{g/p}(x)$ is the unpolarized collinear gluon distribution and
\begin{equation}
{\cal N}_g(x) = N_g x^{\alpha}(1-x)^{\beta}\,
\frac{(\alpha+\beta)^{(\alpha+\beta)}}
{\alpha^{\alpha}\beta^{\beta}}\;\;\;\;\;\; h(k_\perp) = \sqrt{2e}\,\frac{k_\perp}{M'}\,e^{-k_\perp^2/M'^2}\,,
\label{eq:nq-coll}
\end{equation}
with $|N_g|\leq 1$. We also define the parameter $ \rho = M'^2/(\langle k_\perp^2 \rangle_g +M'^2)$, with $0 < \rho < 1$.
Notice that the collinear component of all TMDs defined above evolve through a DGLAP evolution.

In the sequel, by a combined analysis of SSAs for mid-rapidity pion and $D$-meson production, we will show how 
one can constrain the two GSFs in the CGI-GPM. 
Predictions for $p^\uparrow p\to J/\psi\,X$ 
and $p^\uparrow p \to \gamma\, X$ are also given, as well as the corresponding results within the GPM.

\section{Results and comparison with data}

In the GPM approach, via 
the analysis of the SSA in $pp\to\pi^0 X$ at mid-rapidity~\cite{DAlesio:2015fwo}, one can directly extract the GSF
and then give predictions to be compared against other sets of data. 
The parameters obtained, adopting for $\langle k_\perp^2\rangle_g = 1$ GeV$^2$, are~\cite{DAlesio:2018rnv}:
\begin{equation}
\label{eq:par_gsf_GPM}
N_g=0.25\,, \hspace*{1cm} \alpha = 0.6\,, \hspace*{1cm}\beta = 0.6\,, \hspace*{1cm}\rho = 0.1  \,.
\end{equation}
The analysis within the CGI-GPM, for which we have two independent GSFs, appears more difficult. 
The first issue we address is the effective role of the $f$- and $d$-type contributions. 
To do this, we maximize the effects of the two functions, by saturating the positivity bound 
for their $x$-dependent parts (i.e.~${\cal N}_g(x)=\pm 1$) and adopting the value $\rho = 2/3$.

\begin{figure}[t]
\begin{center}
\includegraphics[trim = 1.cm 0cm 1cm 0cm, width=7cm]{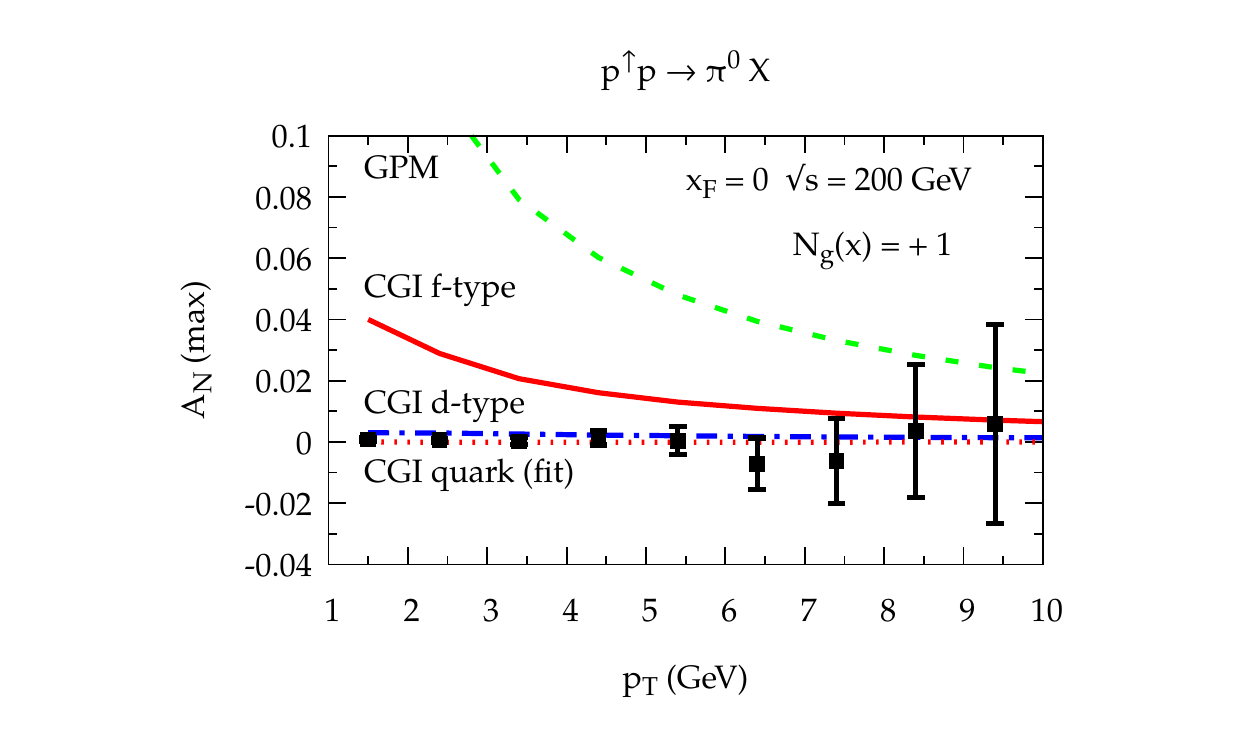}
\includegraphics[trim = 1.cm 0cm 1cm 0cm, width=7cm]{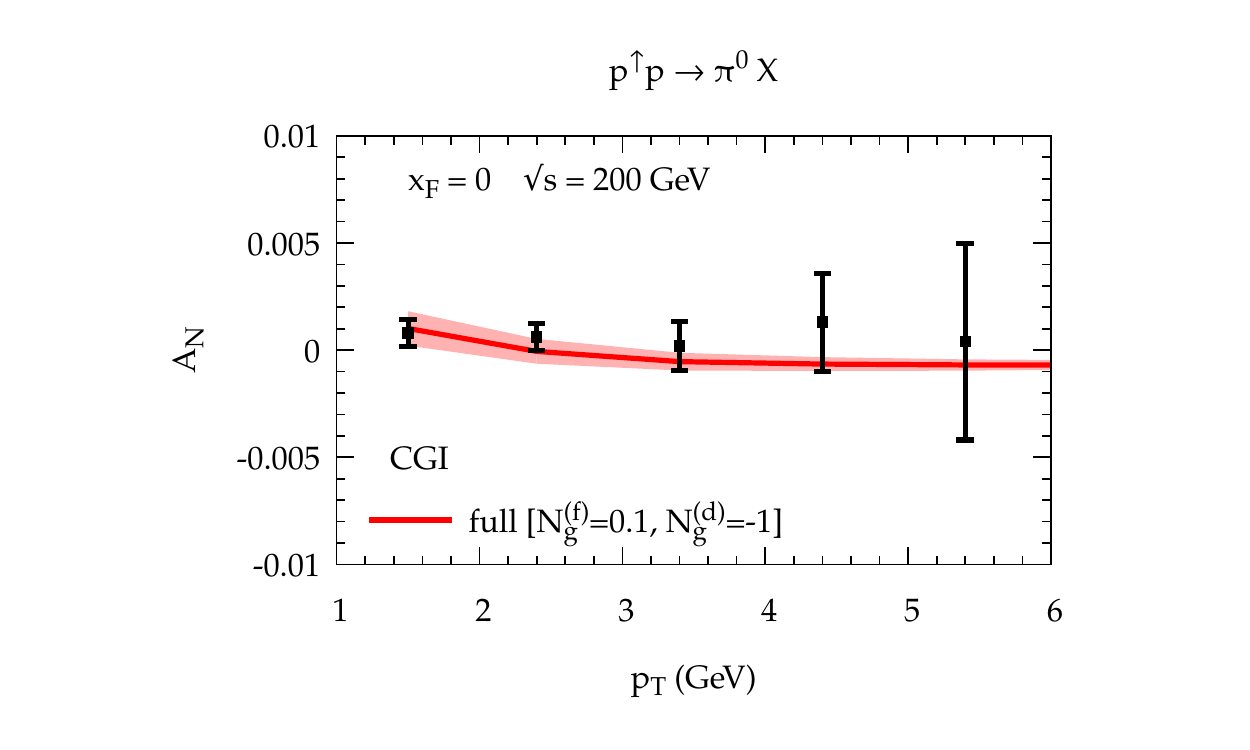}
\caption{Left panel: maximized gluon Sivers contributions (${\cal N}_g(x) =+1$) to $A_N$ for 
$p^\uparrow p\to \pi^0\, X$ as a function of $p_T$. The quark Sivers effect, as extracted from SIDIS fits, 
is also shown (red dotted line). Right panel: $A_N$ estimates obtained adopting 
a reduced $f$-type GSF ((${\cal N}_g^{(f)}(x) =0.1$)) and a negative saturated $d$-type GSF (${\cal N}_g^{(d)}(x) =-1$).
Shaded area represents a $\pm 20$\% uncertainty on ${\cal N}_g^{(f)}$. Data are from Ref.~\cite{Adare:2013ekj}.}
\label{fig:AN-pi}
\end{center}
\end{figure}

In Fig.~\ref{fig:AN-pi} (left panel) we present the maximized (${\cal N}_g(x) = +1$) gluon Sivers 
contributions to $A_N$ for the process $p^\uparrow p\to \pi^0\, X$ at $\sqrt s=200$ GeV 
and mid-rapidity as a function of $p_T$, together with the PHENIX data~\cite{Adare:2013ekj}: 
$f$-type (red solid line), $d$-type (blue dot-dashed line) and GPM (green dashed line).
The quark Sivers contribution (red dotted line) is totally negligible.
More importantly, the $d$-type term appears strongly suppressed. 
This is due to the cancellation between the $gq$ and $g\bar q$ channels, entering with a relative opposite sign,
and the absence of the $gg\to gg$ contribution, 
relevant at moderate values of $p_T$. In contrast, the GPM approach gives the largest maximized effect, 
since its partonic contributions are all positive and unsuppressed.

To obtain SSAs comparable with the very tiny and \emph{positive} data values, 
one would conservatively expect a relative cancellation between the $f$- and $d$-type pieces, 
with a strongly reduced and \emph{positive} $f$-type GSF.
The results, for ${\cal N}_g^{(f)}(x)=+0.1$ and ${\cal N}_g^{(d)}(x)=-1$, are 
shown in the right panel of Fig.~\ref{fig:AN-pi}, 
together with an overall uncertainty band of about $\pm$20\% on ${\cal N}_g^{(f)}$.

\begin{figure}[t]
\begin{center}
\includegraphics[trim = 1.cm 0cm 1cm 0cm, width=7cm]{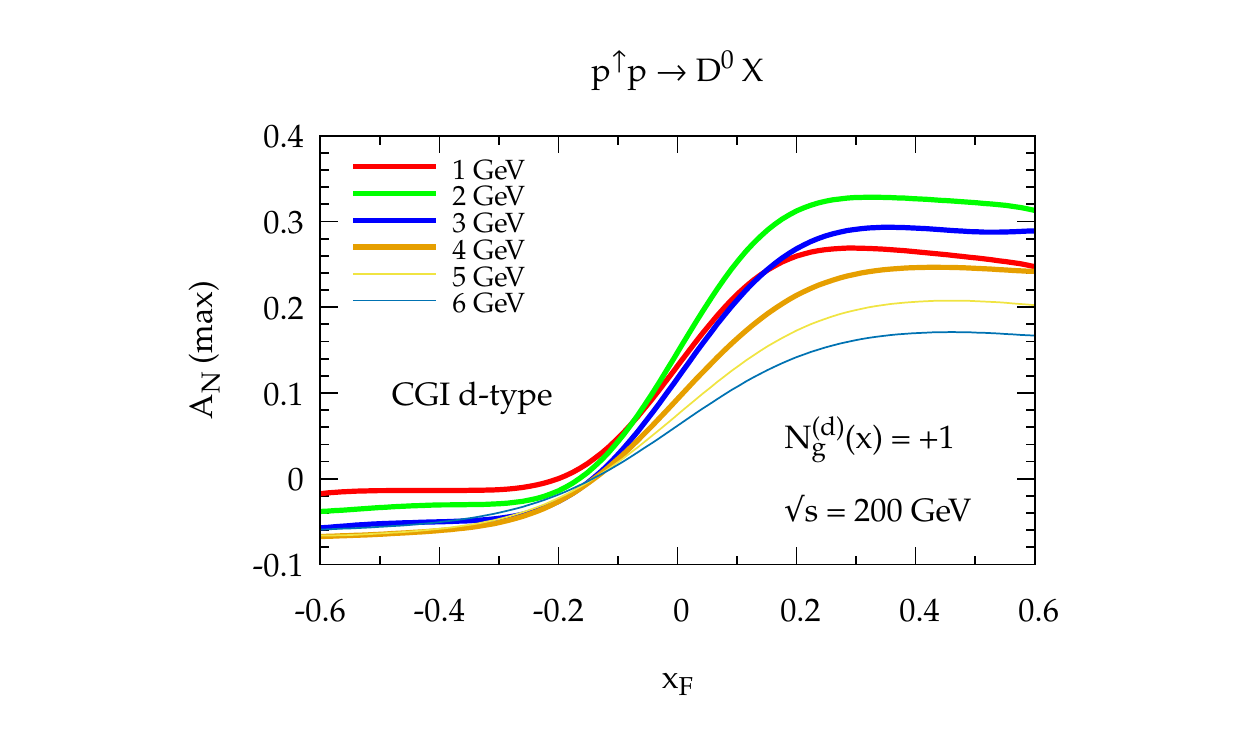}
\includegraphics[trim = 1.cm 0cm 1cm 0cm,width=7cm]{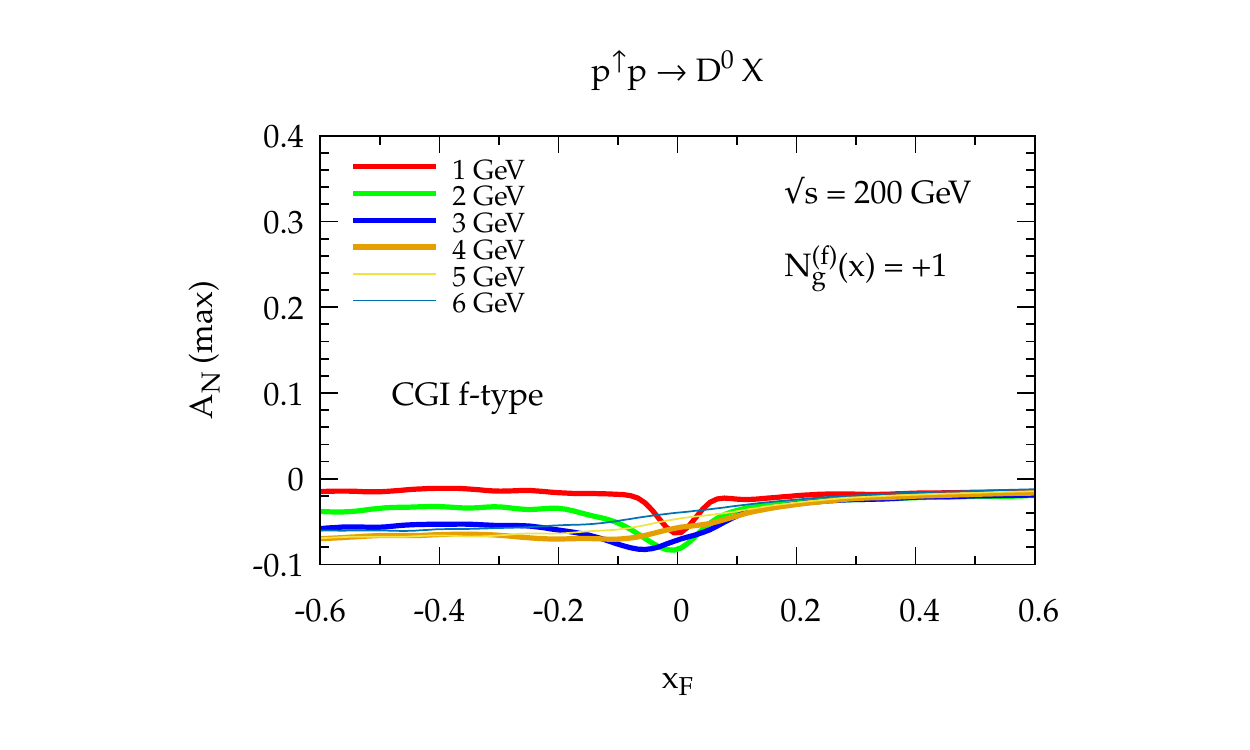}
\caption{Maximized (${\cal N}_g(x) =+1$) $A_N$ for $p^\uparrow p\to D^0\, X$ at different $p_T$ 
values (1 -- 6 GeV) as a function of $x_F$, within the CGI-GPM approach: 
$d$-type (left panel) and $f$-type (right panel) contributions.}
\label{fig:AN-D0-sat}
\end{center}
\end{figure}

Let us now consider $A_N$ for $D^0$ production at $\sqrt s= 200$ GeV 
and the corresponding analysis for its muon decays, 
for which data are available~\cite{Aidala:2017pum}. 
In Fig.~\ref{fig:AN-D0-sat} we show the results for $A_N$ as a function of $x_F$ for different $p_T$ values, 
obtained by separately maximizing the $d$- (left panel) and $f$-type (right panel) contributions. 
In this case, in the forward region the $d$-type term is sizeable, while the $f$-type is suppressed. 
The reason is that, for $D^0$ production, at leading order only the charm-quark 
fragmentation function enters and no cancellation, as in the previous case, occurs. 
Moreover, the hard partonic parts favor the $d$-type w.r.t.~the $f$-type term at $x_F>0$, while 
for $x_F<0$ both effects are almost washed out by the integration over the Sivers azimuthal phase.

Adopting now the results of the analysis of $\pi^0$ SSA data, the $f$-type piece should be accordingly reduced by 
0.1, thus becoming practically negligible.
This means that for $D^0$ production we can safely consider 
only the $d$-type GSF. Similar considerations apply also to $\bar D^0$ production.

\begin{figure}[t]
\begin{center}
\includegraphics[trim = 1.cm 0cm 1cm 0cm, width=7cm]{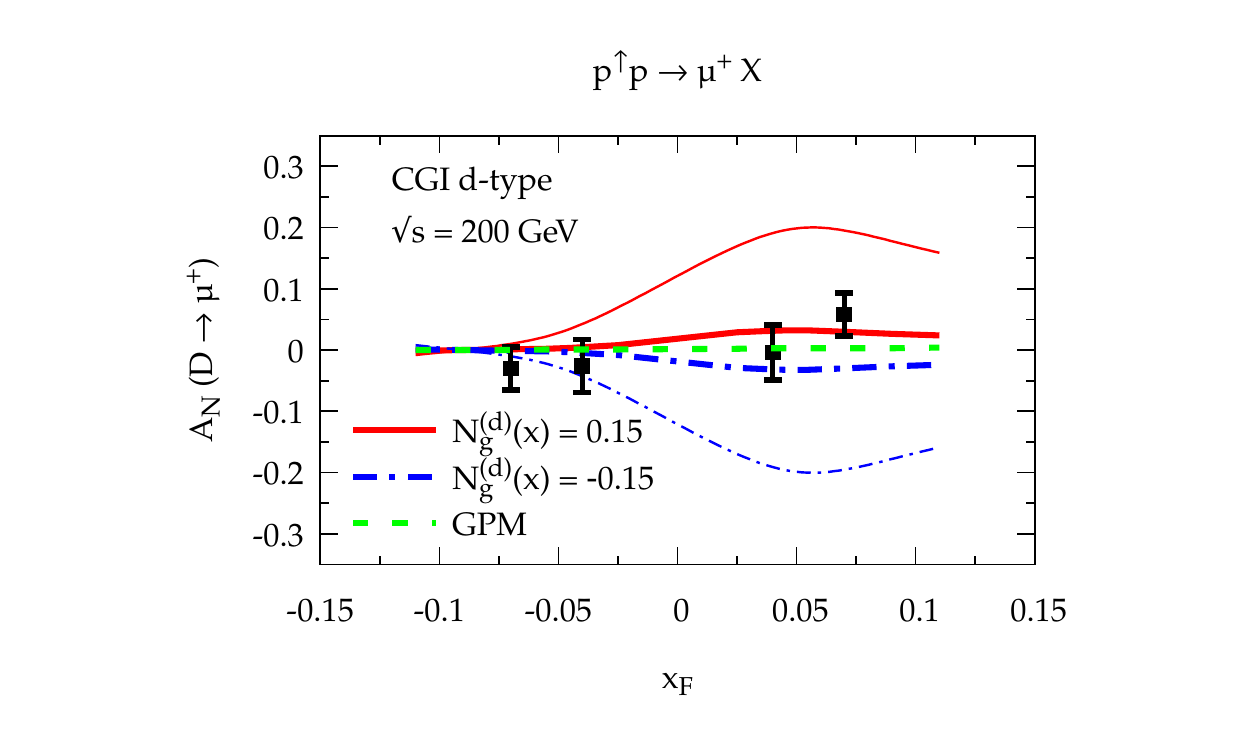}
\includegraphics[trim = 1.cm 0cm 1cm 0cm, width=7cm]{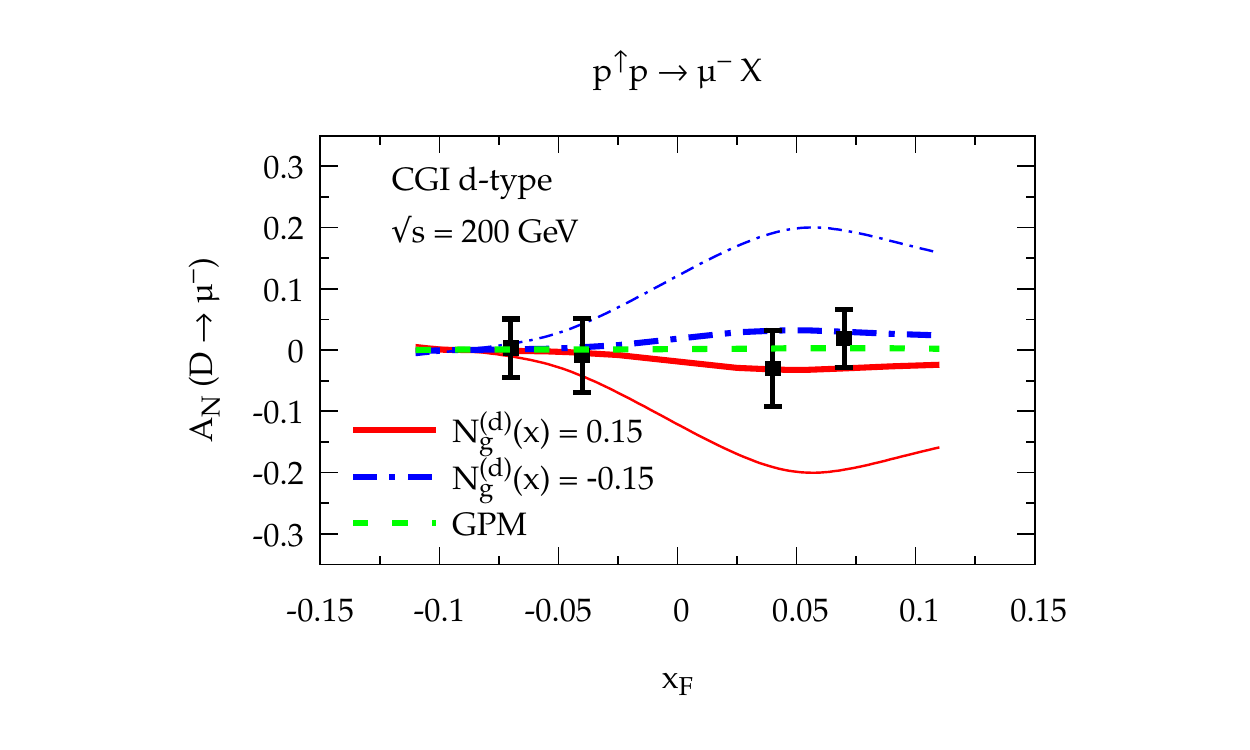}
\caption{$d$-type gluon contributions to $A_N$ for $p^\uparrow p\to \mu^+\, X$ (left panel) and 
$p^\uparrow p\to \mu^-\, X$ (right panel) from $D$-meson decay as a function of $x_F$. Maximized effect:
${\cal N}_g^{(d)}(x) =+1$ (thin red solid lines), ${\cal N}_g^{(d)}(x) =-1$ (thin blue dot-dashed lines).
GPM predictions (green dashed lines) are also shown. Data are from Ref.~\cite{Aidala:2017pum}.}
\label{fig:AN-muon}
\end{center}
\end{figure}

For a direct comparison with data, the estimates for $D$-meson production have 
to be converted to the SSAs for their muon decay products~\cite{Aidala:2017pum}. 
Although the amount and precision of muon data do not allow for a true fit,  
important information can be gathered even adopting a very simple scenario with ${\cal N}_g^{(f,d)}(x)\equiv N_g^{(f,d)}$.

As shown in Fig.~\ref{fig:AN-muon} the maximized $d$-type contributions 
(thin red solid line: ${\cal N}_g^{(d)} =+1$, thin blue dot-dashed line: ${\cal N}_g^{(d)} = -1$) 
largely overestimate the positive $x_F$ data in size. Notice that ${\cal N}_g^{(d)} = -1$, 
together with ${\cal N}_g^{(f)} = + 0.1$, was adopted to reproduce the $\pi^0$ SSA data (see Fig.~\ref{fig:AN-pi}, right panel). 
To get a fair account of the muon SSA data, one can take  $|{\cal N}_g^{(d)}| \leq 0.15$, 
with a mild preference for positive values, because of the positive $\mu^+$ data point. 
In Fig.~\ref{fig:AN-muon}, the results with ${\cal N}_g^{(d)} = +0.15(-0.15)$ are shown 
as thick red solid lines (thick blue dot-dashed lines) 
both for $\mu^+$ (left panel) and $\mu^-$ (right panel) production.

Going back to the pion SSA, we find that by varying ${\cal N}_g^{(d)}$ 
in the range $-0.15\,\div\,+0.15$, while keeping $\rho=2/3$, a very good description of both 
the $\mu^\pm$ and $\pi^0$ data can be obtained if ${\cal N}_g^{(f)}$ 
varies in the \emph{corresponding} range $+0.05\,\div\,-0.01$:
\be
\label{eq:par_gsf_CGI}
{\cal N}_g^{(d)} = - 0.15\,  \rightarrow  {\cal N}_g^{(f)} = +0.05 \hspace*{1cm}
{\cal N}_g^{(d)} = + 0.15\,  \rightarrow  {\cal N}_g^{(f)} = -0.01\,.
\ee
A stronger reduction of the $f$-type GSF is then 
required by the combined analysis of $\pi$ and $\mu$ data.
In contrast, in the GPM approach the parametrization of the GSF 
extracted from the $\pi^0$ SSA data, see Eq.~(\ref{eq:par_gsf_GPM}), 
allows us to describe the SSA for $\mu^\pm$ fairly well (Fig.~\ref{fig:AN-muon}, green dashed lines).

\noindent
Note that a comparable good agreement with data is reached within the twist-three approach~\cite{Koike:2011mb}.

\begin{figure}[b]
\begin{center}
\includegraphics[width=8cm]{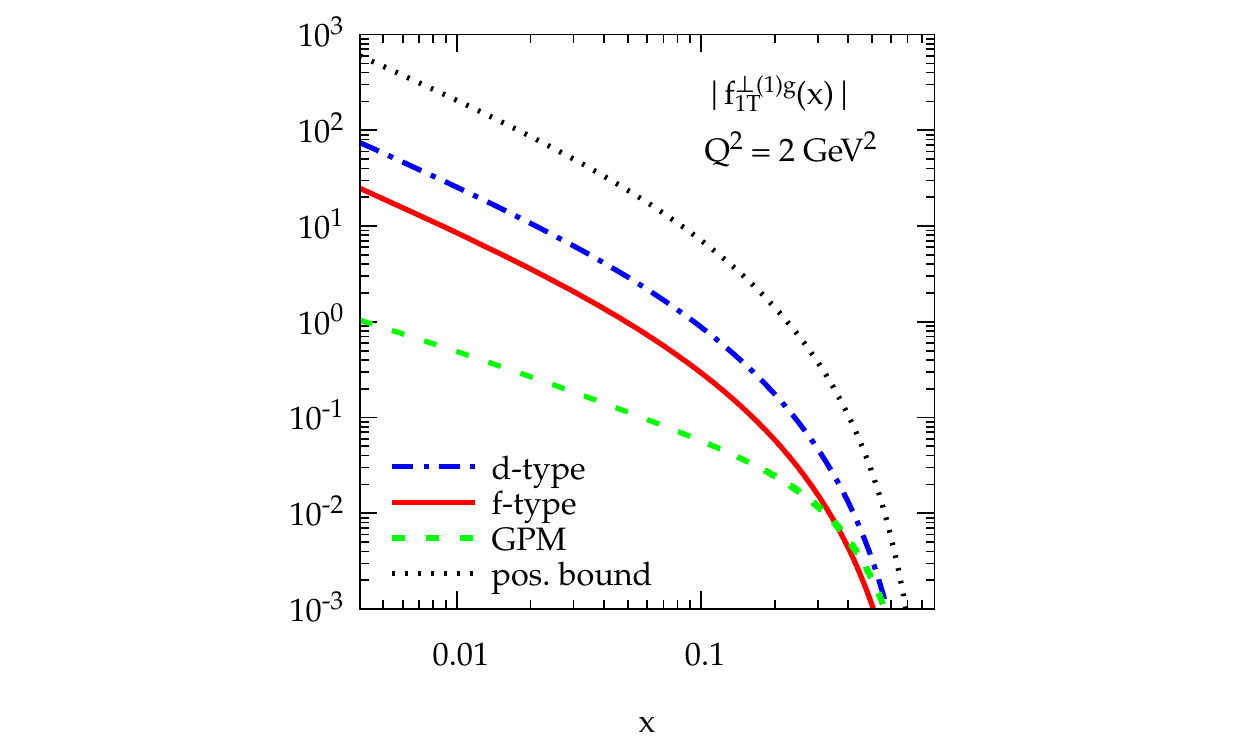}
\caption{Upper values for the first $k_\perp$-moments of the gluon Sivers functions at $Q^2 = 2$ GeV$^2$.
}
\label{fig:1stmom}
\end{center}
\end{figure}

The absolute value of the first $k_\perp$-moment of 
the GSFs is shown in Fig.~\ref{fig:1stmom} for the GPM (green dashed line) 
and the CGI-GPM approaches, $d$-type (blue dot-dashed line) and $f$-type ($N_g^{(f)}=0.05$, red solid line), 
together with the positivity bound (black dotted line).  Notice that
the GSF in the GPM approach, for which the hard partonic parts are all positive, is the smallest.

\section{Predictions}

$A_N$ for $J/\psi$ production is directly sensitive to the gluon Sivers function, 
and, within the CGI-GPM approach and the Color Singlet model, 
only to the $f$-type distribution~\cite{DAlesio:2017rzj}.

\begin{figure}[t]
\begin{center}
\includegraphics[trim = 1.cm 0cm 1cm 0cm,width=7cm]{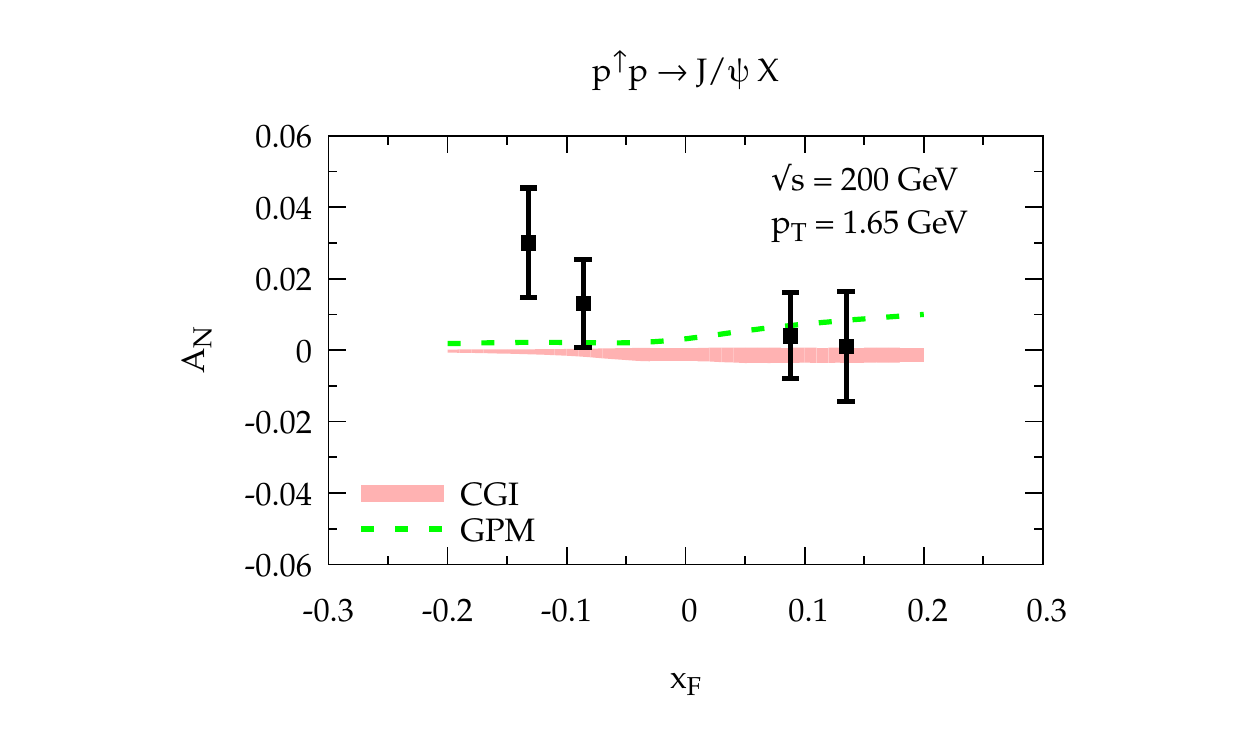}
\includegraphics[trim = 1.cm 0cm 1cm 0cm,width=7cm]{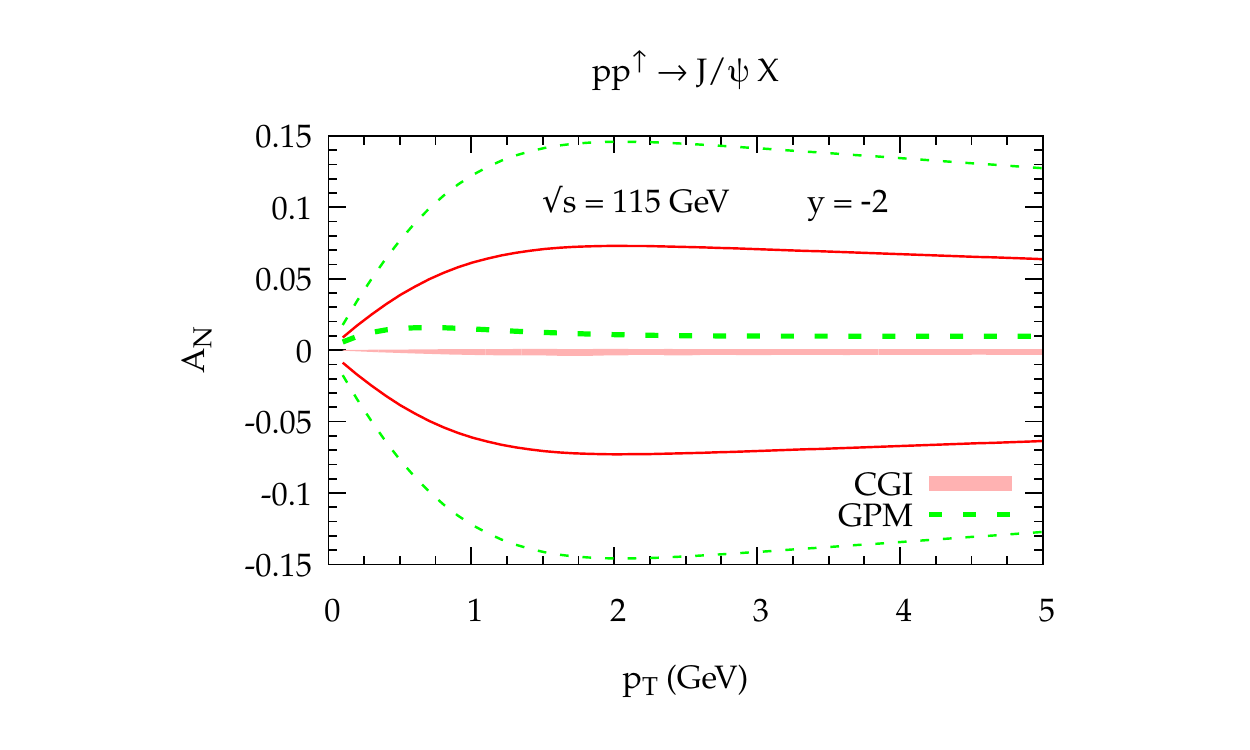}
\caption{Left panel: Predictions of $A_N$ for $p^\uparrow p\to J/\psi\, X$ 
at $\sqrt s=200$ GeV as a function of $x_F$ (RHIC set-up). Data are from Ref.~\cite{Aidala:2018gmp}.
Right panel: $A_N$ for $pp^\uparrow \to J/\psi\, X$ at $\sqrt{s} = 115$ GeV and $y=-2$ 
as a function of $p_T$ (LHC fixed target set-up): predictions (thick lines) and maximized contributions (thin lines). 
}
\label{fig:Jpsi}
\end{center}
\end{figure}

In Fig.~\ref{fig:Jpsi} (left panel) we show our estimates against PHENIX data~\cite{Aidala:2018gmp} 
for $A_N$ in $p^\uparrow p\to J/\psi\, X$: 
GPM (green dashed line), CGI-GPM (red band: $-0.01\le {\cal N}_g^{(f)} \le 0.05$). 
With the exception of one point at $x_F<0$, 
data are compatible with zero and well described by our estimates.

The corresponding analysis of $A_N$ at LHC in the fixed polarized target 
mode ($\sqrt s=115$ GeV) could be also extremely useful.  
Here one could probe even larger $x$ in the polarized proton, 
accessing the gluon TMDs in a complementary region. 
Notice that in such a configuration the backward rapidity region refers to the forward region 
for the polarized proton target. 
Predictions within the GPM (thick green dashed lines) and the CGI-GPM (red bands) approaches, 
together with the corresponding upper/lower positivity bounds (thin lines) 
are shown in the right panel of Fig.~\ref{fig:Jpsi}.

\begin{figure}[t]
\begin{center}
\includegraphics[trim = 1.cm 0cm 1cm 0cm,width=7cm]{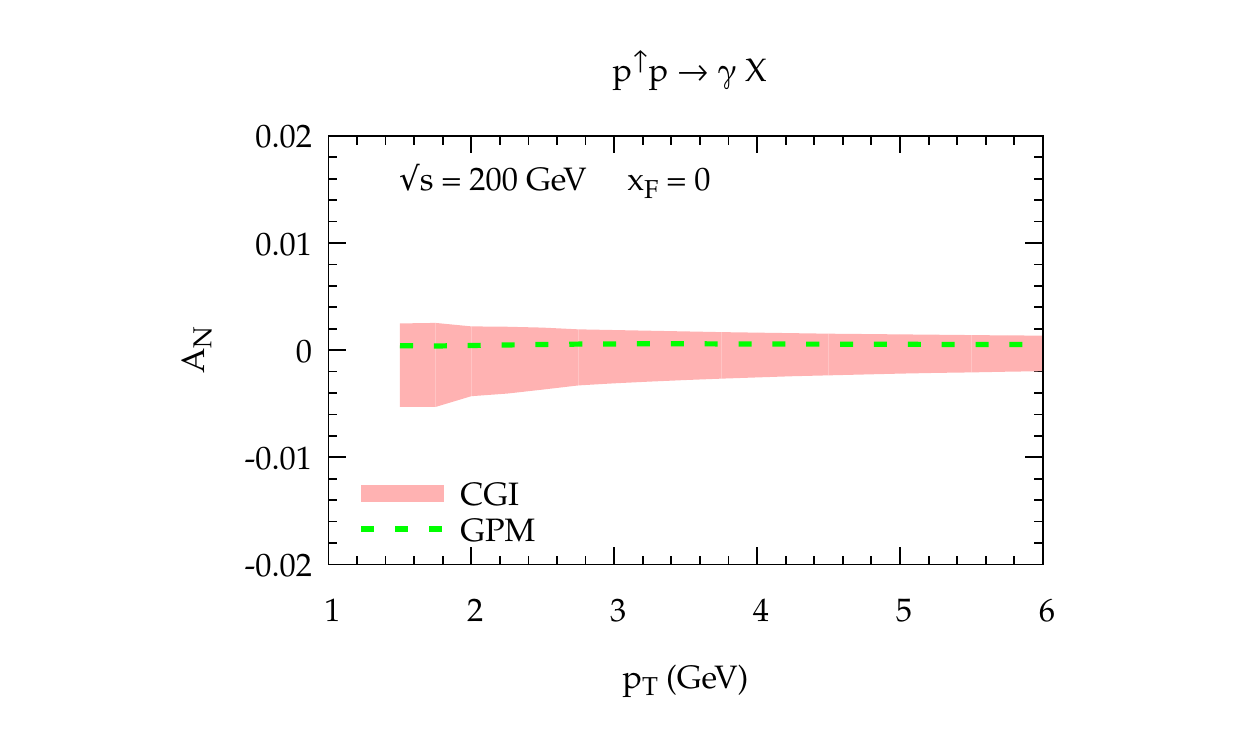}
\includegraphics[trim = 1.cm 0cm 1cm 0cm,width=7cm]{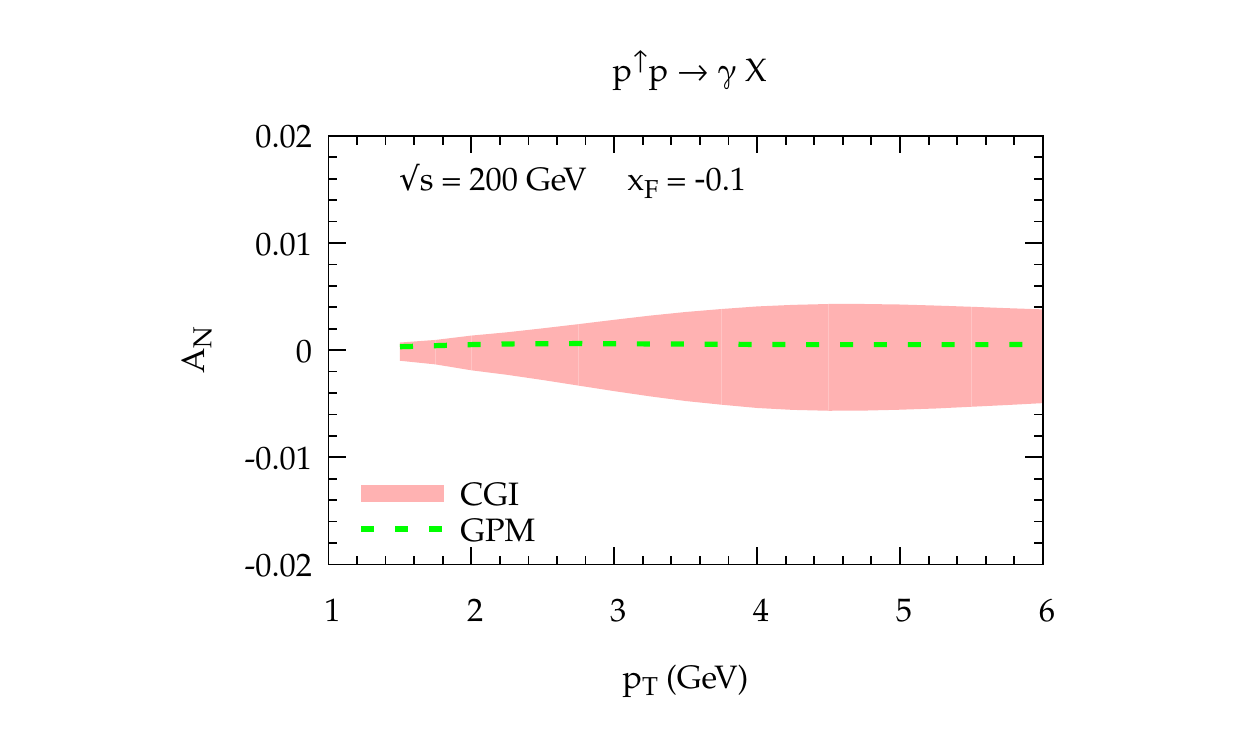}
\caption{Predictions of $A_N$ for $p^\uparrow p\to \gamma\, X$  
as a function of $p_T$ (see Eqs.~(\ref{eq:par_gsf_GPM}), (\ref{eq:par_gsf_CGI})).
}
\label{fig:AN-gamma}
\end{center}
\end{figure}

SSAs for direct photon production is another ideal tool to access the GSF.
In Fig.~\ref{fig:AN-gamma} we show our predictions obtained adopting Eqs.~(\ref{eq:par_gsf_GPM}), (\ref{eq:par_gsf_CGI}).  
Despite of their tiny values, a measure of $A_N$ for direct photon production 
would be extremely important to test the consistency of this picture.

Summarizing, the present analysis can be considered the first detailed attempt 
towards a quantitative extraction of the process dependent GSFs in the CGI-GPM scheme. 
Even if it is not yet possible to clearly discriminate between the GPM and the CGI-GPM approaches, the results are encouraging. 
More data, and more precise, will certainly help in shedding light on this important TMD, 
as well as on the validity of the whole approach.


\providecommand{\href}[2]{#2}\begingroup\raggedright\endgroup

\end{document}